\documentclass[prl,amsmath,amssymb,twocolumn]{revtex4-2}
\usepackage{graphicx}
\usepackage{subfigure}
\usepackage{adjustbox}
\usepackage{bm}
\usepackage{color}
\usepackage{braket}
\usepackage{standalone}
\usepackage{multirow}
\usepackage{tikz}
\usepackage{mathrsfs}
\usepackage{dsfont}
\usepackage{comment}
\usepackage{amsmath}
\usepackage[colorlinks,bookmarks=true,citecolor=blue,linkcolor=red,urlcolor=blue]{hyperref}
\usepackage{cleveref}

\newcommand{\elliptic}[5][\scriptstyle]{\vartheta\left[\begin{array}{c}{{#1 #2}}\\{#1 #3}\end{array}\right]\left(#4\middle|#5\right)}

\newcommand{\Pf}{{\rm Pf}}
\newcommand{\PfTilde}{{\rm \widetilde{Pf}}}

\graphicspath{{./figures/}}

\begin{document}

\title{Signatures of Supersymmetry in the $\nu{=}5/2$ Fractional Quantum Hall Effect}
\author{Songyang Pu$^1$, Ajit C. Balram$^{2,3}$, Mikael Fremling$^{4}$,  Andrey Gromov$^{5}$, and Zlatko Papi\'c$^{1}$}

\affiliation{$^{1}$School of Physics and Astronomy, University of Leeds, Leeds LS2 9JT, United Kingdom}
\affiliation{$^{2}$Institute of Mathematical Sciences, CIT Campus, Chennai 600113, India}
\affiliation{$^{3}$Homi Bhabha National Institute, Training School Complex, Anushaktinagar, Mumbai 400094, India} 
\affiliation{$^{4}$Institute for Theoretical Physics and Center for Extreme Matter and Emergent Phenomena, Utrecht University, Princetonplein 5, 3584 CC Utrecht, Netherlands}
\affiliation{$^5$Department of Physics \& Condensed Matter Theory Center, University of Maryland, College Park, Maryland 20740, USA}

\date{\today}

\begin{abstract}
The Moore-Read state, one of the leading candidates for describing the fractional quantum Hall effect at filling factor $\nu{=}5/2$, is a paradigmatic $p$-wave superconductor with non-Abelian topological order. Among its many exotic properties, the state hosts two collective modes: a bosonic density wave and a neutral fermion mode that arises from an unpaired electron in the condensate. It has recently been proposed that the descriptions of the two modes can be unified by postulating supersymmetry (SUSY) that relates them in the long-wavelength limit. Here we extend the SUSY description to construct wave functions of the two modes on closed surfaces, such as the sphere and torus, and we test the resulting states in large-scale numerical simulations. We demonstrate the equivalence in the long-wavelength limit between SUSY wave functions and previous descriptions of collective modes based on the Girvin-MacDonald-Platzman ansatz, Jack polynomials, and bipartite composite fermions. Leveraging the first-quantized form of the SUSY wave functions, we study their energies using the Monte Carlo method and show that realistic $\nu{=}5/2$ systems are close to the putative SUSY point, where the two collective modes become degenerate in energy. 

\end{abstract}

\maketitle

{\bf \em Introduction.---}Exotic topological properties of fractional quantum Hall (FQH) fluids, such as quantized Hall resistance~\cite{Tsui82}, and excitations with fractional charge~\cite{Laughlin83, de-Picciotto97} and fractional statistics~\cite{Leinaas77, Wilczek82, Arovas84, Nakamura20, Bartolomei20}, have been the subject of major research efforts over the past decades. More recently, FQH states have come into renewed focus due to their unique \emph{geometric} properties such as the Hall viscosity~\cite{Avron95, Read09, Haldane09, Pu20} and the Girvin-MacDonald-Platzman (GMP) magnetoroton mode~\cite{Girvin85, Girvin86}. The latter is a low-lying collective mode of any gapped FQH fluid and corresponds to a bosonic excitation that can be viewed as a density wave, analogous to the roton in superfluid $^{4}$He~\cite{Feynman1953}. The GMP mode has been observed in several experiments using inelastic light scattering and surface acoustic waves~\cite{Pinczuk93, Kang01, Kukushkin09}.

\begin{figure}
    \centering    
    \includegraphics[width=\linewidth]{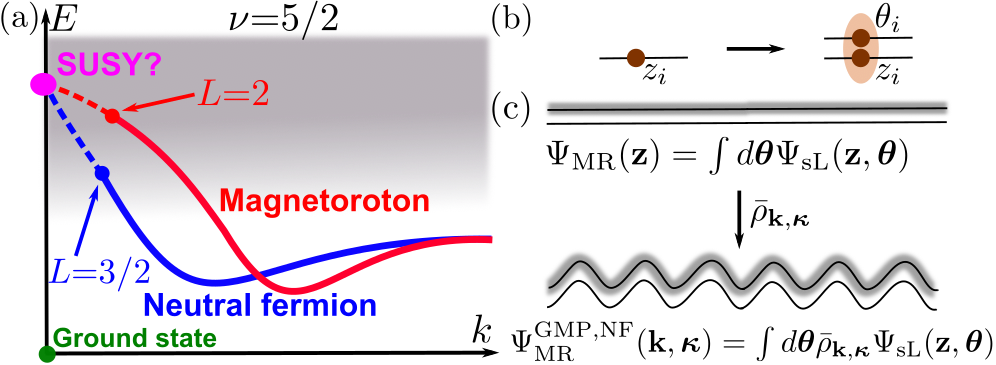}
    \caption{
    (a) Sketch of the energy spectrum of a gapped $\nu{=}5/2$ FQH state with two neutral collective modes, the magnetoroton and neutral fermion. In the long-wavelength limit $k\ell{\to}0$, the two modes carry angular momenta $L{=}2$ and $L{=}3/2$, respectively, and their degeneracy would give rise to an emergent SUSY. (b) Both modes can be compactly described by introducing, for each particle's coordinate $z_i$, its SUSY partner with a Grassmann coordinate $\theta_i$~\cite{Gromov20}. (c) When the total number of particles is even, the Moore-Read state is recovered by integrating out $\boldsymbol{\theta}$ variables (depicted by shading) from $\Psi_{sL}$, the Laughlin wave function at $\nu=1/2$ in the superplane~\cite{Hasebe2008}. Similarly, the wave functions for the GMP and NF modes are obtained by acting with the projected density operator, $\bar{\rho}_{\mathbf{k},\boldsymbol{\kappa}}$, which generates a density wave in superspace. 
    }
    \label{fig:summary}
\end{figure}

However, certain FQH states, such as the one observed at filling fraction $\nu{=}5/2$~\cite{Willett87}, can possess an \emph{additional} collective mode, suggested by the numerical simulations~\cite{Bonderson11b, Moller11, Papic12} [see Fig.~\ref{fig:summary}(a) for a schematic illustration]. The additional collective mode in the $\nu{=}5/2$ state is naturally accounted for by the Moore-Read (MR) wave function~\cite{Moore91} and its particle-hole conjugate the anti-Pfaffian state~\cite{Levin07, Lee07}, two of the leading candidates for understanding the incompressible state at half filling of the second Landau level (LL). The MR state represents a $p$-wave superconductor~\cite{Read96, Read00} of composite fermions, i.e., the bound states of electrons and vortices~\cite{Jain89}. Hence, via analogy with Bardeen–Cooper–Schrieffer (BCS) superconductors~\cite{DeGennes99}, the MR state possesses a fermionic collective excitation -- the ``neutral fermion" (NF) -- which is the analog of the Bogoliubov-de Gennes quasiparticle~\cite{Greiter91, Read00}. The applications of the MR state in topological quantum computation~\cite{Nayak08} rest critically on the gap of the NF mode, as the latter can be excited in the process of `fusion' of two elementary excitations of the MR state that simulate the action of quantum gates~\cite{Nayak96}.

In the long-wavelength limit, corresponding to momenta $k{\ll}\ell^{{-}1}$, where $\ell$ is the magnetic length, there is a sharp distinction between the bosonic GMP and fermionic NF modes: the former carries integral angular momentum $L{=}2$ on the sphere~\cite{Bonderson11, Sreejith11}, while the latter carries \emph{half odd-integral} $L{=}3/2$~\cite{Moller11, Yang12b, Sreejith11b}. While the GMP mode has a simple description in terms of acting the lowest LL (LLL) projected density operator on the ground state~\cite{Girvin85}, the microscopic description of the NF mode has required far more elaborate constructions~\cite{Sreejith11b, Rodriguez12b, Yang12b, Yang13a, Repellin15} that are not amenable to numerical studies in large systems. Recently, Ref.~\cite{Gromov20} proposed a unified description of the GMP and NF modes in the MR state based on the two modes being supersymmetry (SUSY) partners, see Fig.~\ref{fig:summary}(b)-(c) [see also Ref.~\cite{Ma2021} for the consequences of SUSY for the edge physics of the MR state]. While this provides an elegant description of both modes via generalization of the GMP ansatz to superspace, the resulting description has not been tested in numerics. Moreover, it remains unclear if the physical $\nu{=}5/2$ system satisfies the assumption of an emergent SUSY.

In this paper, we extend the construction of Ref.~\cite{Gromov20} to formulate fully antisymmetric, LLL-projected wave functions for both the GMP and NF modes adapted to closed manifolds, such as the sphere and torus, that can be evaluated for large system sizes. We numerically demonstrate the equivalence of these wave functions in the long-wavelength limit to alternative descriptions of the modes based on Jack polynomials~\cite{Yang12b} and bipartite composite fermions~\cite{Sreejith11, Sreejith11b, Rodriguez12b}.  We test the existence of SUSY \emph{a posteriori}, by evaluating the energies of the two collective modes in the long-wavelength limit using the Monte Carlo method. While the pure Coulomb interaction at $\nu{=}5/2$ gives rise to a weak breaking of SUSY, with the SUSY gap being about $20\%$ of the excitation gap, tuning the interaction by slightly enhancing the $V_1$ Haldane pseudopotential~\cite{Haldane83, Prange87} leads to the restoration of SUSY. Importantly, the two modes become degenerate in the long-wavelength limit at approximately the same value of $V_1$ that maximizes the ground state overlap with the MR wave function~\cite{Rezayi00}. Our results suggest that the SUSY structure is intrinsically present in spectral properties of the $\nu{=}5/2$ state, allowing us to probe the conditions for its emergence in large-scale numerics. Moreover, our work provides a foundation for developing an effective field theory of paired FQH states that incorporates SUSY as an emergent symmetry of infrared physics.

{\bf \em SUSY wave functions on the sphere.---}We first construct the SUSY wave functions in the spherical geometry~\cite{Haldane83} [see Supplemental Material (SM)~\cite{SOM} for the corresponding construction on the torus]. We assume there are $N$ electrons confined to the surface of a sphere, with a Dirac monopole at the center, emanating a radial magnetic flux of strength $2Qhc/e$. The radius of the sphere is $R{=}\sqrt{Q}\ell$. The total angular momentum $L$ and its $z$-component $M$ are good quantum numbers. The magnitude of the planar wavevector $k$ is given by $k{=}L/R$. 
In terms of the spinor coordinates $u_j$, $v_j$ of the $j$th electron~\cite{Haldane83}, the MR wave function~\cite{Moore91} on the sphere is
\begin{equation}
\label{MR sphere}
\Psi_{\rm MR}=\Pf \left({1\over u_iv_j-u_jv_i} \right) \Phi_1^2,
\end{equation}
where Pf($A$) stands for the Pfaffian of a skew-symmetric matrix $A$ and $\Phi_1 {=} \prod_{i{<}j}\left(u_iv_j{-}u_jv_i\right)$ denotes the Laughlin-Jastrow factor~\cite{Prange87}.

A neutral density wave excitation of the MR state, carrying angular momentum $L$, is described by the GMP ansatz~\cite{Girvin85},
$
\Psi_{\rm GMP}(L){=}\mathcal{P}_{\rm LLL}\hat \rho_{L,M}\Psi_{\rm MR}
$.
Here $\hat \rho_{L, M}{=}\sum_{i{=}1}^N Y_{L, M}(\theta_i,\phi_i)$ is the density operator in the spherical geometry, expressed in terms of spherical harmonics $Y_{L, M}$~\cite{He94}, and $\mathcal{P}_{\rm LLL}$ is the LLL projection operator.  Since the states with different $M$ are degenerate due to rotational symmetry on the sphere, we can set $M{=}{-}L$ for simplicity, in which case (up to normalization) we have $Y_{L,-L}\left(u_j,v_j\right) {\propto} v_j^L\bar{u}_j^L$, where the bar denotes complex conjugation. We see that $\Psi_{\rm GMP}(L)$ needs to be explicitly projected to the LLL and this must be done carefully to maintain the efficiency of its evaluation via Monte Carlo simulations. Due to this obstacle, previous approaches~\cite{Park99b} could only access the dispersion of the GMP mode via the static structure factor of the ground state. 

\begin{figure}[t]
		\includegraphics[width=\linewidth]{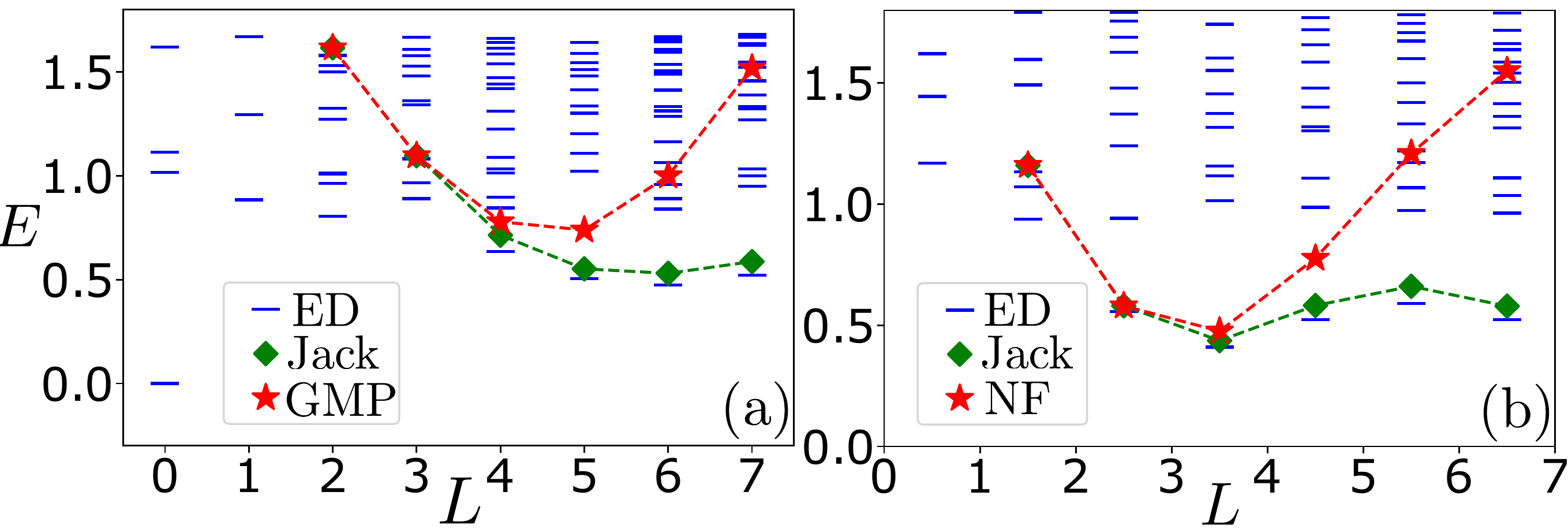}
		\caption{
  The energy spectrum of the three-body parent Hamiltonian of the Moore-Read state on the sphere~\cite{Greiter91} for $N{=}14$ (a) and $N{=}13$ electrons (b). Blue dashes are energy levels obtained by exact diagonalizations (ED), green diamonds are the energies of collective modes constructed via Jack polynomials~\cite{Yang12b}, while red stars are the energies of the GMP wave function in (a) and the NF wave function in (b). At small $L$, both collective modes merge with the continuum of the spectrum. The Jack construction captures the entire collective modes, while the GMP and NF wave functions are accurate only in the long wavelength limit. 
		}
  \label{exact}
\end{figure}

\begin{figure*}
		\includegraphics[width=\textwidth]{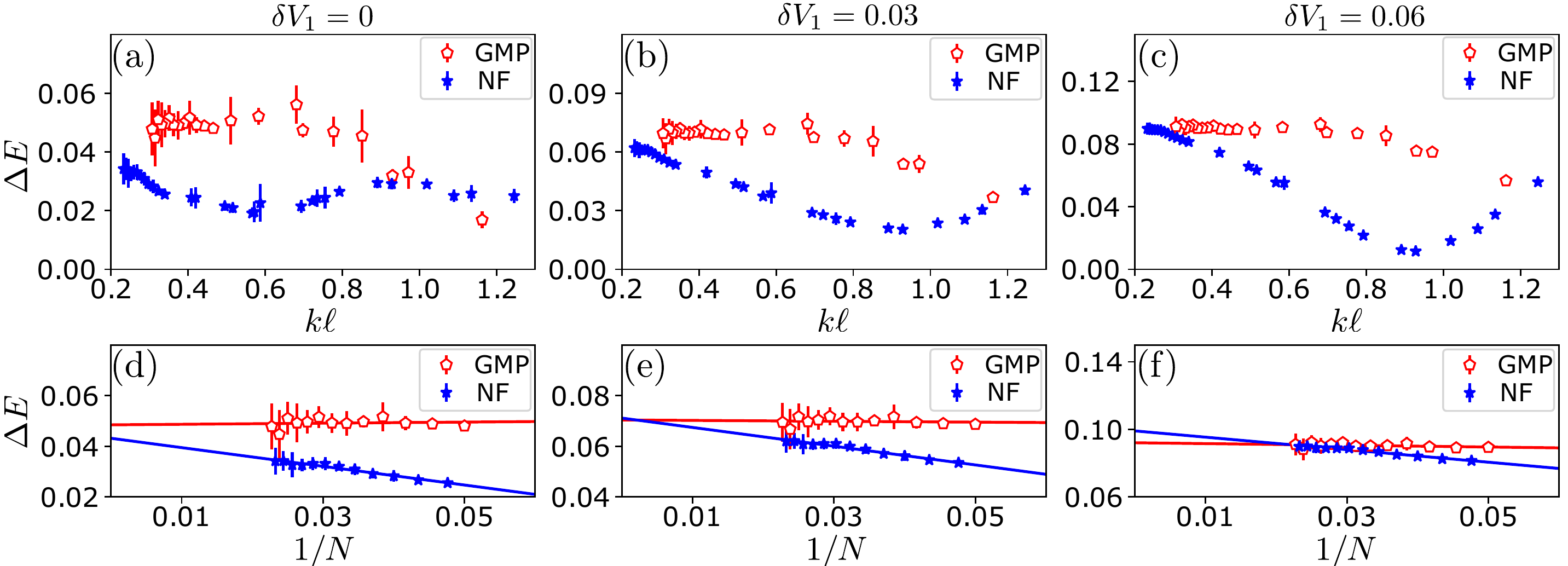}
		\caption{Dispersion of the GMP and NF modes (a)-(c) and the finite-size scaling of the energies of $L{=}2$ and $L{=}3/2$ states (d)-(f). The data is for $20{\leq }N{\leq}44$ electrons for the Coulomb interaction projected to the second LL, with added pseudopotential $\delta V_1$ [(a) and (d) are for pure Coulomb interaction ($\delta V_1 {=} 0$), (b) and (e) are for $\delta V_1{=}0.03$, (c) and (f) are for $\delta V_1{=}0.06$]. The energies are quoted relative to the ground state energy, i.e., $\Delta E{=}E{-}E_0$, in units of $e^2/\epsilon\ell$. Here $E_0$ is the energy of the MR state at the same electron number for the GMP mode, while for the NF mode $E_0$ is the average value of the MR state energy for $N{+}1$ and $N{-}1$ electrons.
		}
  \label{extrapolation}
\end{figure*}

Remarkably, we find that the LLL projection of the GMP mode of the MR state can be performed \emph{exactly}, with the resulting wave function given by 
\begin{align}
\label{GMP1}
\Psi_{\rm GMP}(L) = \sum_{m=1}^N v_m^L \Phi_1^2 \; \Pf \left( {\tilde{U}^{L}_{m,i,j}\over u_iv_j-u_jv_i}\right),
\end{align}
where $\tilde{U}^{L}_{m, i, j}$ is a function of all the electrons' coordinates, defined through a recursion relation (see SM~\cite{SOM}). The derivation of Eq.~\eqref{GMP1} is inspired by the Jain-Kamilla method of projecting composite fermion wave functions~\cite{Jain97, Jain97b}. However, while the latter method yields a wave function different from that obtained from direct projection, we emphasize that the projection in Eq.~\eqref{GMP1} results in exactly the same wave function as the one obtained from direct projection. This form allows us to efficiently compute properties of GMP states in large systems, without relying on the static structure factor.

Our exact projection method furthermore allows extending the NF wave function, formulated as a SUSY partner of the GMP mode in the infinite plane~\cite{Gromov20}, to the spherical geometry. The proposed NF wave function from Ref.~\cite{Gromov20}, after stereographic mapping~\cite{Fano86}, is given by 
\begin{align}
\label{SUSY2}
\nonumber \Psi_{\rm NF}(L) &= \mathcal{P}_{\rm LLL} \Big[\sum_m (-1)^m{1\over u_m}\PfTilde_m\left({1\over u_iv_j-u_jv_i}\right) \\ & Y_{L,-L}\left(u_m,v_m\right) \Phi_1^2 \Big],  
\end{align}
where $\PfTilde_m$ denotes the Pfaffian of the same matrix as in Eq.~\eqref{MR sphere} but with $m$th electron unpaired, i.e., $i, j{\neq}m$, assuming $N$ is odd.  

The wave function in Eq.~\eqref{SUSY2} is clearly not analytic due to ${1/u_m}$ factor. To fix this issue, we replace $1/u_m {\to}{\bar{u}_m}$ in Eq.~\eqref{SUSY2}, which preserves the total magnetic flux and the shift~\cite{Wen92}, resulting in a wave function that is an eigenstate of total angular momentum $L^{2}$. 
Following a similar procedure to the GMP mode, we obtain the NF wave function projected to the LLL:
\begin{align}
\label{SUSY4}
\Psi_{\rm NF}(L)=&\sum_m (-1)^m v_m^L  \Phi_1^2 \; \PfTilde_m\left({1\over u_iv_j-u_jv_i}\right)  P_m(L{+}1),
\end{align}
where $P_m$ is defined through a recursion relation in~\cite{SOM}.

{\bf \em Testing SUSY wave functions.---}To assess the accuracy of the constructed SUSY-based wave functions, in Fig.~\ref{exact} we compare their energies with the exact spectrum of the three-body parent Hamiltonian of the MR state~\cite{Greiter91}, as well as against the modes obtained from Jack polynomials~\cite{Yang12b}. The three-body energies of the GMP and NF wave functions are evaluated in Fock space. 
The GMP wave function can be directly obtained in Fock space by applying the LLL-projected density operator $\hat \rho_{L, M}$ on the MR state. To obtain the NF wave function in Fock space, we use the method of Refs.~\cite{Sreejith11, Balram20a} which involves evaluating all the relevant $L^{2}$ eigenstates and expanding Eq.~\eqref{SUSY4} in that basis. 

First, we note that both the $L{=}1$ GMP and $L{=}1/2$ NF wave functions vanish with machine precision accuracy, consistent with previous findings~\cite{Sreejith11, Sreejith11b, Rodriguez12b, Yang12b}.
Second, the long-wavelength part of the dispersion shows excellent agreement between the GMP (NF) states and the Jack polynomial states from Ref.~\cite{Yang12b}. In fact, by numerically evaluating the overlap between the Jack states and the SUSY states above, we find the Jack states are \emph{identical} to the GMP states in Eq.~\eqref{GMP1} at $L{=}2, 3$ for even $N{\leq}14$, while the Jack states and NF states in Eq.~\eqref{SUSY4} are identical at $L{=}3/2, 5/2$ for odd $N{\leq}13$. Furthermore, all these states appear to be identical to bipartite composite fermion states from Refs.~\cite{Sreejith11, Sreejith11b}, as they give the same overlap with eigenstates of the MR three-body interaction and the Coulomb interaction at $\nu{=}5/2$ (up to the second digit quoted in Ref.~\cite{Sreejith11}). At higher values of $L$, we expect all schemes to produce distinct states. For example, the overlap between the Jack and SUSY-based states decreases at higher values of $L$~\cite{SOM}, and the GMP and NF wave functions no longer capture the exact dispersion at finite momenta $k\ell {\gtrsim} 1$ (i.e., $L {\gtrsim} N$). In the following, we focus on the long-wavelength part of the dispersion, where all the constructions result in the same state, as shown above, allowing us to put variational bounds on the conditions for SUSY to emerge. 

{\bf \em Emergence of SUSY in realistic systems.---}The SUSY construction assumes that the GMP and NF modes have the same energy in the long-wavelength limit. Using the wave functions constructed above, we can probe the SUSY degeneracy in the limit $k\ell{\to}0$ by evaluating the energies via Monte Carlo on large spheres with $R{\gg}\ell$\cite{SOM}. 
The dispersions of the GMP and NF modes are shown in Fig.~\ref{extrapolation}(a)-(c) for the Coulomb interaction projected to the second LL. Furthermore, we explore the neighborhood of the Coulomb interaction by adding a small amount of $\delta V_1$ pseudopotential~\cite{Haldane83}, which captures some of the modifications to the interaction potential due to the finite width of the sample and LL mixing. 

Fig.~\ref{extrapolation}(a) shows that the energy of the GMP mode for the second LL Coulomb interaction is higher than that of the NF mode. However, the GMP dispersion is fairly flat while the energy of the NF mode increases as $k$ decreases. This allows for the possibility of these two modes to meet at $k{=}0$. To check this, we performed a finite-size extrapolation of the energies of $L{=}2$ GMP and $L{=}3/2$ NF states as a function of $1/N$ in Fig.~\ref{extrapolation}(d). The GMP energy is slightly higher than the NF energy in the thermodynamic limit, by $0.005(2)e^2/\epsilon\ell$ which is about 20\% of the $\nu{=}5/2$ excitation gap \cite{Morf02}.

The addition of a small amount of $\delta V_1{\approx} 0.03$ significantly alters the shape of the modes: it makes the GMP mode flatter and the NF mode rise faster at small $k$, see Fig.~\ref{extrapolation}(b). This completely suppresses the gap between the GMP mode and NF mode in the $k\ell{\rightarrow} 0$ limit, as shown in Fig.~\ref{extrapolation}(e). Note that around the same value $\delta V_1{\approx} 0.03$, the overlap of the exact ground state with the MR state is maximized~\cite{Rezayi00, Storni10}. Thus, we conclude that adding a small amount of $V_1$ pseudopotential -- approximately $10\%$ of its value in the second LL -- enhances the overlap with the MR state while at the same time gives rise to the SUSY-degeneracy. Finally, upon an even further increase in $\delta V_1$, shown in  Fig.~\ref{extrapolation}(c)-(f), the gap becomes negative, corresponding to the GMP mode being lower in energy than the NF mode. This suggests there is a finite range of $\delta V_1$ where SUSY degeneracy can be observed.

\begin{figure}[tb]
		\includegraphics[width=0.95\linewidth]{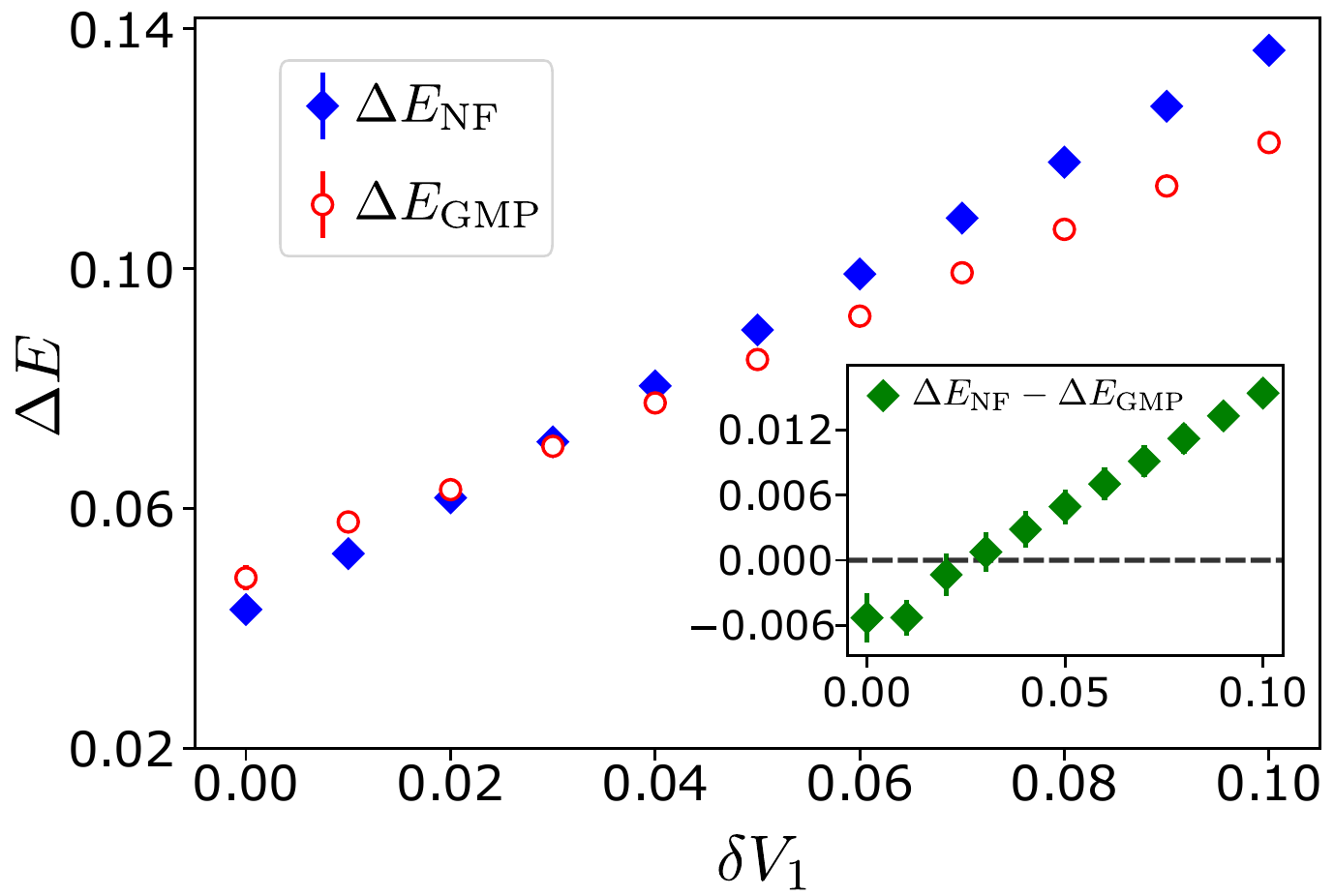}
		\caption{The extrapolated gaps of the $L{=}2$ GMP mode and $L{=}3/2$ NF mode in $N{\to}\infty$ limit, for different values of $\delta V_1$ pseudopotential added to the second LL Coulomb interaction. The inset shows the difference between the gaps. The error only includes the regression error from finite-size extrapolations while ignoring the small statistical error of Monte Carlo. 
		}
  \label{gap}
\end{figure}

A more systematic analysis of the effect of $\delta V_1$ is presented in Fig.~\ref{gap}, which shows the extrapolated gaps of the $L{=}2$ GMP and the $L{=}3/2$ NF states, as a function of $\delta V_1$. The gaps vary linearly with $\delta V_1$, and the slope of the NF gap is larger than that of the GMP gap. This indicates that the NF mode is more sensitive to the hard-core interaction than the GMP mode. We show the difference between the GMP and NF gaps in the long-wavelength limit in the inset of Fig.~\ref{gap}. As a result of the linearity of the GMP and NF energies, their difference is also linear. By varying $\delta V_1$, the difference can be tuned to zero and this happens near $\delta V_1{\approx}0.03$. Increasing $\delta V_1$ beyond this value breaks the SUSY degeneracy. However, in the limit of large $\delta V_1$, we no longer expect our wave functions to provide a good description of the underlying physics, because the effective interaction becomes increasingly LLL-like, which ultimately induces a transition to the gapless composite fermion Fermi liquid~\cite{Halperin93, Rezayi00}.

{\bf \em Conclusions.---}In this work, we have constructed the wave functions for the neutral collective modes of the MR state in the spherical and torus geometries based on the SUSY description. We have shown that, in the long-wavelength limit, these wave functions encompass the previous independent constructions, with the advantage of allowing numerical calculations in large systems. The variational energies of these wave functions suggest that the $\nu{=}5/2$ FQH state is close to the SUSY point and can be driven to it by adding $\delta V_1$. We have confirmed that similar conclusions apply to other modifications of the interaction, e.g., via longer-range pseudopotentials or by increasing the width of the sample, and they also hold for the MR state of bosons at $\nu{=}1$~\cite{SOM}.

The LL-projected Coulomb interaction is particle-hole symmetric, hence our conclusions apply to both the MR and anti-Pfaffian states.
One obvious question is the effect of particle-hole symmetry-breaking interactions on SUSY, such as the three-body interaction that distinguishes the MR and anti-Pfaffian states.
Unfortunately, this question is challenging to address via the method presented here, due to a lack of efficient ways of evaluating three-body interactions in real space. 
A natural source of three-body interactions is LL mixing, which we have neglected above. It would be interesting to explore its effect on the collective modes in order to obtain a closer comparison with experiments.

On the theory side, our demonstration of SUSY at the microscopic level calls for a development of an effective field theory \cite{Gromov17,golkar2016higher, nguyen2018fractional} that incorporates SUSY and treats the two collective modes on the same footing. Beyond the two collective modes discussed here, it would be interesting to understand if SUSY leaves an imprint on the rest of the spectrum of the $\nu{=}5/2$ state, e.g., its dynamical response functions. For example, the weak breaking of SUSY is expected to give rise to a gapless Goldstino mode in the bulk~\cite{SalgadoRebolledo2022}, which could be probed in numerics. Other paired states such as Haldane-Rezayi~\cite{Haldane88a, Nguyen22a}, the Halperin 331 state~\cite{Halperin83}, the $\mathbb{Z}_{n}$ superconductor of composite bosons~\cite{Balram20}, and the ``permanent state"~\cite{Moore91, Green02} are all expected to carry a neutral excitation in addition to the GMP mode, hence it would be interesting to understand if SUSY could emerge in them. Finally, on the experimental front, it would be important to develop protocols for detecting the NF mode, which is invisible to conventional optical probes~\cite{Wurstbauer13}. 

\begin{acknowledgments}
{\bf \em Acknowledgments.---}S.P. and Z.P. acknowledge support by the Leverhulme Trust Research Leadership Award RL-2019-015. A.C.B. and Z.P. thank the Royal Society International Exchanges Award IES$\backslash$R2$\backslash$202052 for funding support.
A.G. was supported in part by NSF CAREER Award DMR-2045181, Sloan Foundation and the Laboratory for Physical Sciences through the Condensed Matter Theory Center.
A.C.B. thanks the Science and Engineering Research Board (SERB) of the Department of Science and Technology (DST) for funding support via the Start-up Grant No. SRG/2020/000154.
M.F. thanks the D-ITP consortium, a program of the Netherlands Organisation for Scientific Research (NWO) that is funded by the Dutch
Ministry of Education, Culture and Science (OCW).
This work was undertaken on ARC3 and ARC4, part of the High Performance Computing facilities at the University of Leeds, UK, and on the Nandadevi supercomputer, which is maintained and supported by the Institute of Mathematical Science's High-Performance Computing Center, India. Some of the numerical calculations were performed using the DiagHam libraries~\cite{diagham}.
\end{acknowledgments}

\clearpage 
\pagebreak

\onecolumngrid
\begin{center}
\textbf{\large Supplemental Online Material for ``Signatures of Supersymmetry in the $\nu{=}5/2$ Fractional Quantum Hall Effect" }\\[5pt]
\vspace{0.1cm}
\begin{quote}
{\small In this Supplementary Material, we first present details of the lowest Landau level (LLL) projection of the SUSY wave functions for the Girvin-MacDonald-Platzman (GMP) and neutral fermion (NF) modes in the spherical geometry. Second, we show how to extend these SUSY wave functions to the torus geometry. Third, we provide details of the effective interaction used in our Monte Carlo calculations. Fourth, we show the dispersion of the two modes and the extrapolation of the long-wavelength limit of the gap between them, termed the ``SUSY gap", for the bosonic Moore-Read (MR) state at $\nu{=}1$. Fifth, we study the effect of varying the $V_3$ pseudopotential and the width of the electron gas on the dispersions of the collective modes at $\nu{=}5/2$.} Finally, we summarize some important details of the Monte Carlo samplings. \\[20pt]
\end{quote}
\end{center}
\setcounter{equation}{0}
\setcounter{figure}{0}
\setcounter{table}{0}
\setcounter{page}{1}
\setcounter{section}{0}
\makeatletter
\renewcommand{\theequation}{S\arabic{equation}}
\renewcommand{\thefigure}{S\arabic{figure}}
\renewcommand{\thesection}{S\Roman{section}}
\renewcommand{\thepage}{S\arabic{page}}
\renewcommand{\thetable}{S\arabic{table}}

\vspace{0cm}

\section{Projected GMP and NF wave functions on the sphere}
\label{LLL}

In this section, we show how to project the GMP and NF wave functions to the LLL. The GMP wave function is
\begin{align}
\label{GMP1app}
\notag \Psi_{\rm GMP}(L) &=  \mathcal{P}_{\rm LLL}\sum_mY_{L,-L}\left(u_m,v_m\right)\Pf \left( {1\over u_iv_j-u_jv_i}\right)  \prod_{i<j}\left(u_iv_j-u_jv_i\right)^2\\ \nonumber
&= \mathcal{P}_{\rm LLL}\sum_mv_m^L\left(\bar{u}_m\right)^{L}\Pf \left( {1\over u_iv_j-u_jv_i}\right) \prod_{i<j}\left(u_iv_j-u_jv_i\right)^2\\
&= \sum_mv_m^L\left({\partial\over\partial u_m}\right)^{L}\Pf \left( {1\over u_iv_j-u_jv_i}\right) \prod_{i<j}\left(u_iv_j-u_jv_i\right)^2.
\end{align}
In going from the second to the third line, we made the replacement $\bar{u}_m{\rightarrow}\partial/ \partial u_m$ which exactly projects the wave function to LLL~\cite{Girvin84b}. However, we still need to move the partial derivative $(\partial/\partial u_m)^{L}$ into the Pfaffian matrix element to enable the numerical calculation for large system sizes. We do this by making use of the fact that $(\partial/\partial u_m)^{L}$ only acts on the coordinate $u_m$, which leads us to
\begin{align}
\Psi_{\rm GMP}(L) =\sum_mv_m^L\prod_{i<j}\left(u_iv_j-u_jv_i\right)^2\Pf \left( {\tilde{U}^{L}_{m,i,j}\over u_iv_j-u_jv_i}\right), 
\end{align}
where we have introduced 
\begin{equation}
\tilde{U}^{L}_{m,i,j}=
\begin{cases}
1, \quad i,j\neq m\\
(u_mv_i-u_iv_m)\prod_{k\neq m}(u_mv_k-u_kv_m)^{-2}\left({\partial\over \partial u_m}\right)^L{\prod_{k\neq m}(u_mv_k-u_kv_m)^{2}\over (u_mv_i-u_iv_m)}, \quad j=m \\
(u_mv_j-u_jv_m)\prod_{k\neq m}(u_mv_k-u_kv_m)^{-2}\left({\partial\over \partial u_m}\right)^L{\prod_{k\neq m}(u_mv_k-u_kv_m)^{2}\over (u_mv_j-u_jv_m)}, \quad i=m\,.
\end{cases}
\end{equation}
To find the explicit form of $\tilde{U}^{L}_{m,i,j}$, we apply the technique developed in the Jain-Kamilla 
 (JK) projection~\cite{Jain97, Jain97b}. First, for $L{=}1$, we find 
\begin{equation}
    \tilde{U}_{m,m,i}^{L=1}=\tilde{U}_{m,i,m}^{L=1}=\sum_{k\neq m,i}{2v_k\over u_mv_k-v_mu_k}+{v_i\over u_mv_i-v_mu_i}.
\end{equation}
For higher $L$, $\tilde{U}^{L}_{m,i,j}$ is evaluated through a recursion relation
\begin{equation}
    \tilde{U}_{m,m,i}^{L+1}=\tilde{U}_{m,i,m}^{L+1}=\left(f_{m,i}(1)+{\partial\over\partial u_m}\right)\tilde{U}_{m,m,i}^{L},
\end{equation}
where $f_{m,i}(L)$ is defined as 
\begin{equation}
    f_{m,i}(L)=2\sum_{k\neq m,i}\left({v_k\over u_mv_k-v_mu_k}\right)^L+\left({v_i\over u_mv_i-v_mu_i}\right)^L.
\end{equation}
This function has the following property under the partial derivative
\begin{equation}
    {\partial \over \partial u_m}f_{m,i}(L)=-Lf_{m,i}(L+1).
\end{equation}
In our numerical calculation, we first obtain the explicit expressions for $\tilde{U}^{L}_{m,i,j}$ for the first few $L$ values using the recursion relation ($L{\leq}6$ in our case) and store these expressions.

The LLL projection of the NF mode is carried out similarly. The NF wave function is
\begin{align}
\label{SUSY3app}
\nonumber \Psi_{\rm NF}(L)=&\mathcal{P}_{\rm LLL}\sum_m (-1)^m{\bar{u}_m}\PfTilde_m \left( {1\over u_iv_j-u_jv_i}\right)  Y_{L,-L}\left(u_m,v_m\right)\prod_{i<j}\left(u_iv_j-u_jv_i\right)^2\\ \nonumber
=&\mathcal{P}_{\rm LLL}\sum_m (-1)^m\PfTilde_m \left( {1\over u_iv_j-u_jv_i}\right) v_m^L\bar{u}_m^{L+1} \prod_{i<j}\left(u_iv_j-u_jv_i\right)^2\\ 
=&\sum_m (-1)^m\PfTilde_m \left( {1\over u_iv_j-u_jv_i}\right) v_m^L\left({\partial\over\partial u_m}\right)^{L+1} \prod_{i<j}\left(u_iv_j-u_jv_i\right)^2.
\end{align}
where ${\PfTilde}_{n}(A_{ij})$ is defined in terms of the antisymmetrizer $\cal A$:
\begin{equation}
 {\PfTilde}_{n}(A_{ij})\equiv \mathcal{A}[A_{12}\cdots A_{n-2,n-1}A_{n+1,n+2}\cdots A_{N-1,N}],
\end{equation}
and in matrix $A$, the $n$th row/column have been removed, hence $N$ is assumed to be odd.

To evaluate how $\left(\partial/\partial u_m\right)^{L{+}1}$ acts on the Jastrow factor $\prod_{i{<}j}\left(u_iv_j{-}u_jv_i\right)^2$, we define a function 
\begin{equation}
 P_m(L)\equiv \prod_{i<j}\left(u_iv_j-u_jv_i\right)^{-2}\left({\partial\over\partial u_m}\right)^{L}\prod_{i<j}\left(u_iv_j-u_jv_i\right)^2,
\end{equation}
such that
\begin{equation}
    \Psi_{\rm NF}(L)=\sum_m (-1)^m\PfTilde_m\left({1\over u_iv_j-u_jv_i}\right) v_m^L\prod_{i<j}\left(u_iv_j-u_jv_i\right)^2P_m(L+1).
\end{equation}
We once again apply the technique of JK projection and obtain the expression for $P_m(L)$ through the recursion relation
\begin{equation}
    P_m(L)=
    \begin{cases}
    1, & L=0\\
    \left(2h_m(1)+{\partial\over \partial u_m}\right)P_m(L-1), & L\geq 1
    \end{cases}
\end{equation}
Here, the function $h_m(L)$ is defined as
\begin{equation}
    h_m(L)=\sum_{j\neq m}\left({v_m\over u_jv_m-v_ju_m}\right)^L.
\end{equation}
It has the following property under the action of derivative
\begin{equation}
    {\partial\over \partial u_m}h_m(L)=-Lh_m(L+1).
\end{equation}
We emphasize that, while we use the technique of JK projection here, our final forms of the LLL projected wave function for both the GMP and NF modes are identical to those obtained from a direct projection. This is in contrast to the composite fermion wave functions obtained by the JK projection, which differ from the composite fermion states obtained by direct projection. The difference originates from the fact that the GMP and NF modes only have a derivative acting on one particle coordinate in each term of the summation and thereby, the process of bringing the derivative operator into the matrix element or Jastrow factor is exact. On the other hand, for the composite fermion states the derivative acts on many particles which leads to differences in the wave functions obtained from the JK method and direct projection.

We test the overlaps between our wave functions and the wave functions constructed from Jack polynomials in Table ~\ref{overlap}. As noted in the main text, the states at the two lowest angular momenta obtained from these different constructions are identical, while the higher angular momentum states are different.

\begin{table}
\begin{tabular}{|c|c|c|c|}
\hline
\multicolumn{2}{|c|}{NF N=13}&\multicolumn{2}{|c|}{GMP N=14}\\ \hline
L&$|\langle \Psi_{\rm NF}|\Psi_{\rm Jack}\rangle|^2$&L&$|\langle \Psi_{\rm GMP}|\Psi_{\rm Jack}\rangle|^2$\\ \hline
3/2&0.99999999999356&2&0.99999999999142\\ \hline
5/2&0.99999999999998&3&0.99999999999732\\ \hline
7/2&0.984873763994&4&0.95789675762548\\ \hline
\end{tabular}
\caption{\label{overlap}The overlap of our NF and GMP wave functions with the corresponding modes constructed from Jack polynomials in Ref.~\cite{Yang12b}. 
We attribute the small deviation from 1 in the first two rows to limited numerical precision in constructing the wave functions (e.g., the construction of Jack states requires a Lanczos diagonalization of $L^2$ operator). 
}
\end{table}

\section{SUSY wave functions of the MR collective modes in the torus geometry}
\label{torus}

The MR wave function in the torus geometry is \cite{Greiter92a, Read96, Chung07}:
\begin{equation}
\label{MR}
\Psi^{(a,b,k_{\rm CM})}_{{\rm MR}} = e^{\sum_i \frac{z_i^2 -|z_i|^2}{4 \ell^2}} \vartheta \begin{bmatrix}
A \\ B
\end{bmatrix}\Bigg({\frac{2Z}{L}} \Bigg |2 \tau \Bigg) 
\Pf\Bigg(\frac{\vartheta \begin{bmatrix}
{a} \\ {b}
\end{bmatrix}
\Bigg( \frac{z_i-z_j}{L}\Bigg | \tau \Bigg )}{\vartheta 
\begin{bmatrix}
\frac{1}{2} \\ \frac{1}{2}
\end{bmatrix}
\Bigg( \frac{z_i-z_j}{L}\Bigg | \tau \Bigg )}\Bigg) \prod_{i<j} \Bigg [ \vartheta 
\begin{bmatrix}
\frac{1}{2} \\ \frac{1}{2}
\end{bmatrix}
\Bigg( \frac{z_i-z_j}{L}\Bigg | \tau \Bigg ) \Bigg ]^2. 
\end{equation}
Here $\elliptic[\displaystyle]abz\tau$ stands for the Jacobi theta function with rational characteristics~\cite{Mumford07}, which is defined as
\begin{equation}
\elliptic[\displaystyle]abz\tau=\sum_{n=-\infty}^{\infty}e^{i\pi \left(n+a\right)^2\tau}e^{i2\pi \left(n+a\right)\left(z+b\right)}.
\end{equation}
The constants 
\begin{eqnarray}
A=\frac{\phi_1}{4\pi }+\frac{k_{\rm CM}}{2}+ \frac{(N-1)}{2}+\frac{(1-2a)}{4}, \quad B=-\frac{\phi_2}{2\pi } + (N-1)-\frac{(1-2b)}{2},
\end{eqnarray}
depend on the center-of-mass degeneracy index $k_{\rm CM}{=}0,1$ and the topological degeneracy indices $(a,b)$. The parameters $(a,b)$ take the values $(0,0)$, $(1/2,0)$, and $(0,1/2)$, which correspond to the topological degeneracy in Haldane momentum sectors $\mathbf{K}_0\equiv(K_x,K_y){=}(N/2,N/2),(0,N/2)$, and $(N/2,0)$. The angles $\phi_j$ are the boundary conditions in the two principal directions of the torus. 

The GMP mode~\cite{Girvin86, Girvin87} is written as
\begin{equation}
 |\Psi_{\rm GMP}(\mathbf{k})\rangle=\mathcal{P}_{\rm LLL}\frac{1}{\sqrt{N}}\sum_{j=1}^Ne^{i\mathbf{k}\cdot \mathbf{r}_j}|\Psi_0\rangle.
\end{equation}
The plane-wave factor $e^{i\mathbf{k}\cdot \mathbf{r}_j}$ after LLL projection becomes 
\begin{equation}
 \mathcal{P}_{\rm LLL}e^{i\mathbf{k}\cdot \mathbf{r}_j}=e^{\frac{i}{2}\bar{k}(z_j+ik)}e^{ik\partial_{z_j}}\mathcal{P}_{\rm LLL}.
\end{equation}
One of the advantages of using the torus geometry is that we can treat the translation operators $T_j(ik){=}e^{ik\partial_{z_j}}$ directly, which is not possible in the spherical geometry. The torus is mapped into periodic parallelograms with two edges $L_1$ and $L_2{=}L_1\tau$. The allowed momenta in the torus geometry which preserves the periodic boundary conditions are $\mathbf{k}{=}n_1\mathbf{b}_1{+}n_2\mathbf{b}_2$, where $\mathbf{b}_1{=}\left(\frac{2\pi}{L_1},{ {-}2\pi {\rm Re}[\tau]\over L_1{\rm Im}[\tau]}\right)$ and $\mathbf{b}_2{=}\left(0,{2\pi\over L_1{\rm Im}[\tau]}\right)$. The $n_1$ and $n_2$ take integer values $0,1,2{\cdots}N_\phi{-}1$. The $(n_1,n_2{+}N)$ density operator acting on $k_{\rm CM}{=}0$ MR state is equivalent to the $(n_1,n_2)$ density operator acting on $k_{\rm CM}{=}1$ MR state, therefore $(n_1,n_2)$ and $(n_1,n_2{+}N)$ have the same energy while the CM momenta are different. On the other hand, $(n_1,n_2)$ and $(n_1{+}N,n_2)$ do not necessarily have identical energy.
The GMP mode of the MR state in the torus geometry can thus be written as
\begin{align}\label{eq:SUSY_GMP_app}
\notag \Psi^{(a,b,k_{\rm CM})}_{{\rm GMP-MR}}\left(k\right)&
=e^{\sum_i \frac{z_i^2 -|z_i|^2}{4 \ell^2}}e^{-{k\over 4}\left(k+2\bar{k}\right)} 
\vartheta \begin{bmatrix} {A} \\ {B} \end{bmatrix}
\Bigg({\frac{2(Z+ik)}{L}} \Bigg |2 \tau \Bigg) \times \\ 
&\sum_{n=1}^NT_n(ik)
\Pf\Bigg(\frac{\vartheta \begin{bmatrix}
{a} \\ {b}
\end{bmatrix}
\Bigg( {z_i-z_j\over L}\Bigg | \tau \Bigg )}{\vartheta 
\begin{bmatrix}
{1\over 2} \\ {1\over 2}
\end{bmatrix}
\Bigg( \frac{z_i-z_j}{L}\Bigg | \tau \Bigg )}\Bigg)\prod_{i<j} \Bigg [ \vartheta 
\begin{bmatrix}
{1\over 2} \\ {1\over 2}
\end{bmatrix}
\Bigg( {z_i-z_j\over L}\Bigg | \tau \Bigg ) \Bigg ]^2. 
\end{align}
The Haldane momentum of $\Psi^{(a,b,k_{\rm CM})}_{{\rm GMP-MR}}\left(k\right)$ is $\mathbf{k}{+}\mathbf{K}_0$, where $\mathbf{K}_0$ is decided by $(a,b)$, as specified above. 

Similarly, we can make the density operator (translation operator) act only on the Jastrow factor, and obtain the NF mode in the torus geometry
\begin{align}
\label{eq:NF_SUSY_app}
\notag \Psi^{(a,b,k_{\rm CM})}_{{\rm SUSY-MR}}\left(k\right)&=
e^{\sum_i \frac{z_i^2 -|z_i|^2}{4 \ell^2}}e^{-{k\over 4}\left(k+2\bar{k}\right)}
\vartheta \begin{bmatrix} {A} \\ {B} \end{bmatrix}
\Bigg({\frac{2(Z+ik)}{L}} \Bigg |2 \tau \Bigg) \times\\ 
&\sum_{n=1}^N (-1)^{n+1} {\PfTilde}_{n}\Bigg(\frac{\vartheta \begin{bmatrix}
{a} \\ {b}
\end{bmatrix}
\Bigg( {z_i-z_j\over L}\Bigg | \tau \Bigg )}{\vartheta 
\begin{bmatrix}
{1\over 2} \\ {1\over 2}
\end{bmatrix}
\Bigg( \frac{z_i-z_j}{L}\Bigg | \tau \Bigg )}\Bigg)T_n(ik)\prod_{i<j} \Bigg [ \vartheta 
\begin{bmatrix}
{1\over 2} \\ {1\over 2}
\end{bmatrix}
\Bigg( {z_i-z_j\over L}\Bigg | \tau \Bigg ) \Bigg ]^2, 
\end{align}
where ${\PfTilde}_{n}(A_{ij})$ is defined in the previous section.
Here since $N$ is odd, both $\mathbf{K}_0$ and $\mathbf{k}$ are half-integers, while the final momentum $\mathbf{K}_0{+}\mathbf{k}$ are integers. 

We note that neutral collective modes of the MR states in the torus geometry have previously been constructed using an antisymmetrized double-layer ansatz with twisted boundary conditions~\cite{Repellin15}. It would be interesting to test and compare this ansatz with the wave functions in Eqs.~\eqref{eq:SUSY_GMP_app}-\eqref{eq:NF_SUSY_app} and, in particular, check whether they may be identical in special momentum sectors. 

\section{Effective interaction}
\label{interaction}

An arbitrary two-body rotationally symmetric interaction in the spherical geometry can be defined through the Haldane pseudopotentials~\cite{Haldane83, Prange87} $V_{m}$ which specify the energy penalty for placing two particles in the relative angular momentum $m$ state. However, it is difficult to directly evaluate Haldane pseudopotentials in Monte Carlo. Instead, one can construct a real-space effective interaction that reproduces the same pseudopotentials as the target interaction. In our calculation, we use the following form of the effective interaction to simulate the second LL (SLL) Coulomb interaction between electrons with some amount of $\delta V_1$ interaction added to it~\cite{Balram13b}
\begin{equation}
V^{\rm eff}(r)={1\over r}+{1\over \sqrt{r^6+1}}+{9\over 4\sqrt{r^{10}+10}}+\left(C_0+C_1r^2+C_2r^4\right)e^{-r^2}.
\label{eq: eff_int_SLL_dV1}
\end{equation}
We fix the values of coefficients $C_0, C_1$ and $C_2$ such that the values of $V_1$, $V_3$, and $V_5$ pseudopotentials of the effective interaction $V^{\rm eff}(r)$ match the corresponding pseudopotentials of the second LL Coulomb interaction with $\delta V_1$ added to it. The numerical values used in our calculation for various $\delta V_1$ are listed in Table~\ref{coff}. Earlier studies have confirmed that the effective interaction is a good approximation to the target interaction since the two interactions only differ in pseudopotentials $V_m$ with $m{\geq}7$. This leads to only minor differences in the total energy of states~\cite{Shi07, Shi08} since the energies are predominantly determined by the small-$m$ pseudopotentials $V_{m}$.

\begin{table}[h]
\begin{tabular}{|c|c|c|c|}
\hline
$\delta V_1$ & $C_0$ & $C_1$ & $C_2$\\ \hline
0.0&-20.94019&12.98214&-1.53215\\ \hline
0.01&-20.19019&12.55245&-1.48332\\ \hline
0.02&-19.44019&12.12276&-1.43449\\ \hline
0.03&-18.69019&11.69308&-1.38566\\ \hline
0.04&-17.94019&11.26339&-1.33684\\ \hline
0.05&-17.19019&10.83370&-1.28801\\ \hline
0.06&-16.44019&10.40401&-1.23918\\ \hline
0.07&-15.69019&9.97433&-1.19035\\ \hline
0.08&-14.94019&9.54464&-1.14152\\ \hline
0.09&-14.19019&9.11495&-1.09270\\ \hline
0.10&-13.44019&8.68526&-1.04387\\ \hline
\end{tabular}
\caption{\label{coff}The coefficients of the effective interaction $V^{\rm eff}$ defined in Eq.~\eqref{eq: eff_int_SLL_dV1} that reproduces the same $V_1$, $V_3$ and $V_5$ pseudopotentials as the second LL Coulomb interaction with some $\delta V_1$ added to it. }
\end{table}

The total energy includes the electron-electron, electron-background, and background-background interactions. The sum of the latter two terms can be evaluated exactly for the SLL Coulomb plus $\delta V_1$ interaction as~\cite{Balram20b}
\begin{equation}
 {1\over \sqrt{Q-1}}+\delta V_1{4Q-1\over (2Q+1)^2},   
\end{equation}
where the flux $2Q{=}2N{-}3$ for the MR state. The gap for the neutral fermion presented in the main text is given with respect to the ground state energy of an odd number of electrons. Since the MR state can only be constructed for an even number of electrons, the ground state energy of an odd number of electrons is obtained from an interpolation of the energies of even number of electrons. The interpolation is carried out on the per-particle energies which include the above background contribution.

\begin{figure*}
		\includegraphics[width=0.31\textwidth]{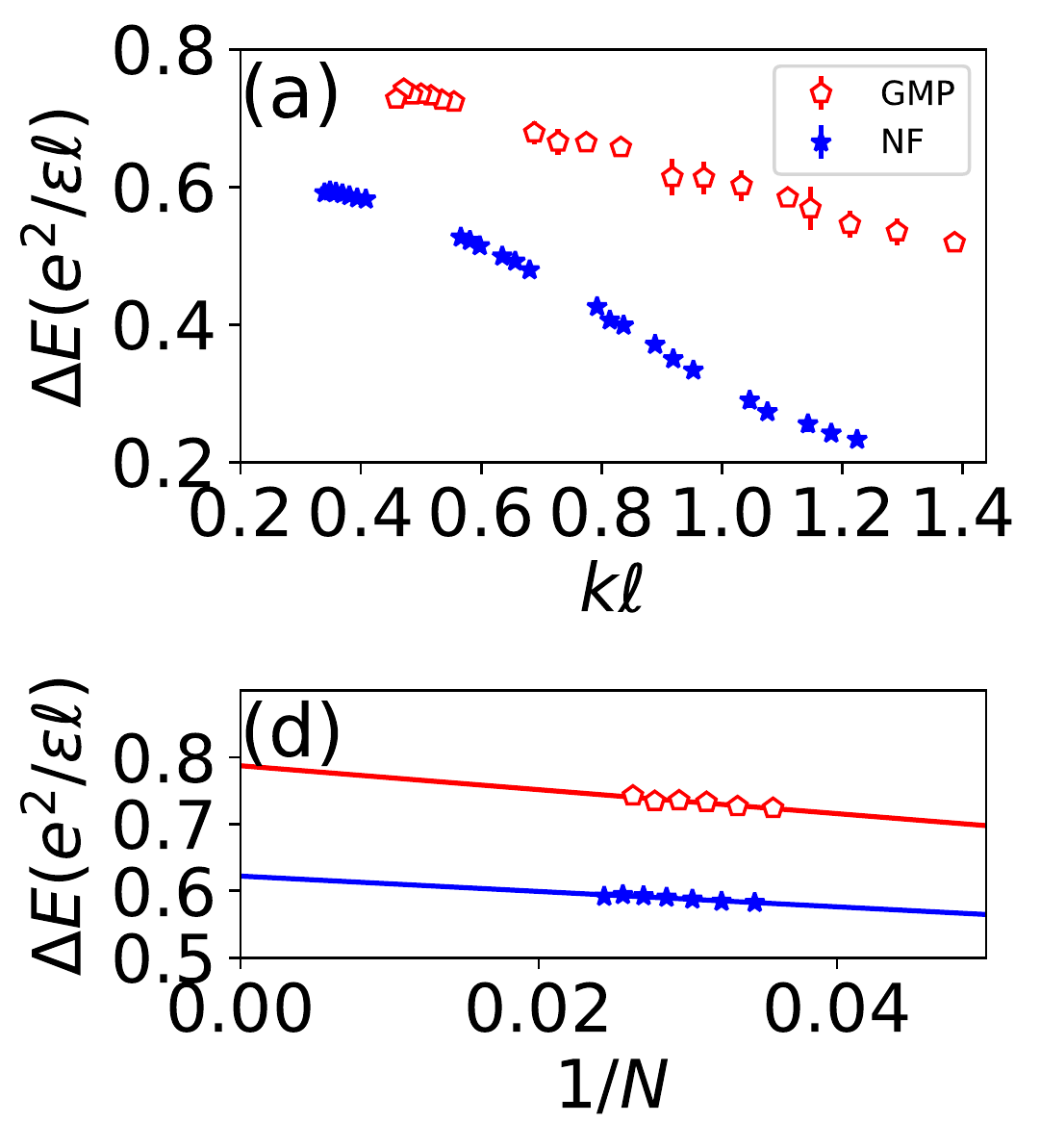}
        \includegraphics[width=0.31\textwidth]{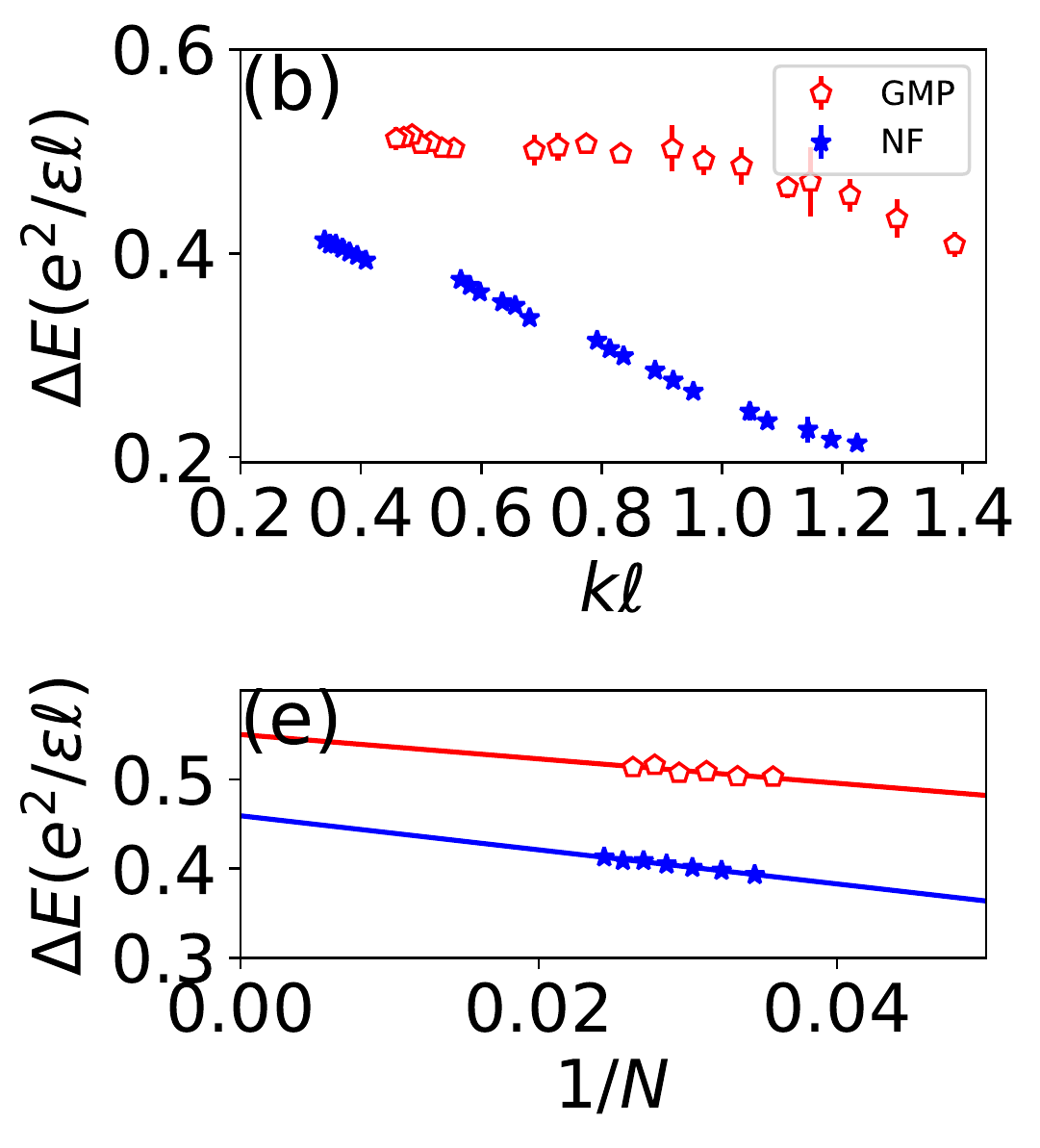}
        \includegraphics[width=0.31\textwidth]{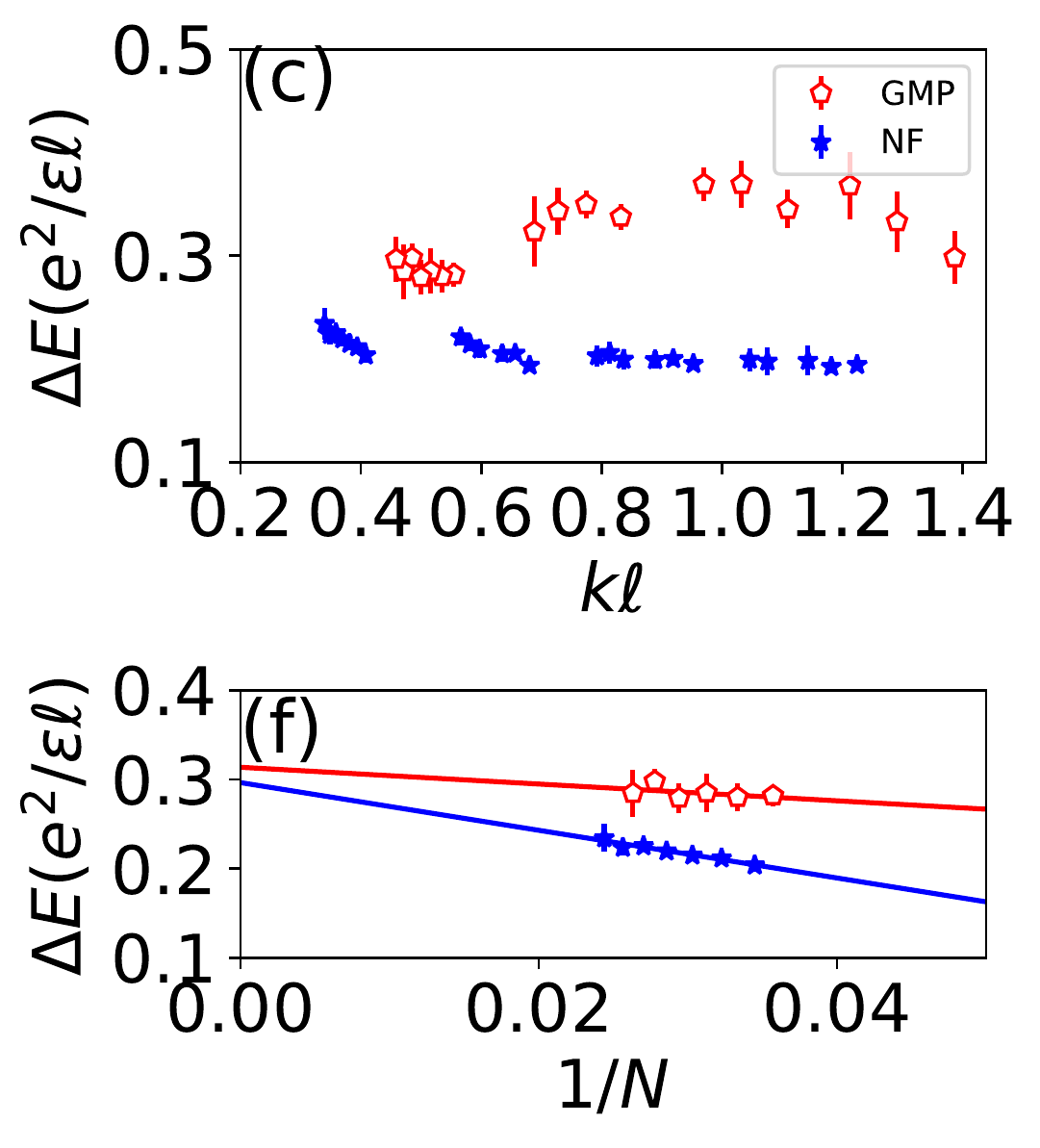}
		\caption{Energy dispersion (a)-(c) and finite-size extrapolation (d)-(f) of the GMP and NF mode energies for bosons at $\nu{=}1$. The interaction is LLL-projected Coulomb with some $\delta V_2$ pseudopotential, where $\delta V_2{=}0$ for (a) and (d), $\delta V_2{=}0.04$ for (b) and (e), $\delta V_2{=}0.08$ for (c) and (f). The energies shown are measured relative to the ground state energy, i.e., $\Delta E{=}E{-}E_0$. The plots include data from various system sizes for $N {\in} [28, 42]$. The finite-size extrapolation uses the energies of $L=2$ GMP mode and $L{=}3/2$ NF mode only. All energies are in units of $e^2/\epsilon\ell$.
		}
  \label{bosonic}
\end{figure*}

\section{Bosonic Moore-Read state at $\nu{=}1$}
\label{bosonic-state}

While in the main text, we focused on the electronic system at $\nu{=}5/2$, the MR state is also known to occur for bosons in the LLL at $\nu{=}1$~\cite{Cooper01, Regnault03}.
The bosonic state is constructed in the same manner as the fermionic state, but with only one factor of $\Phi_1$ instead of two. Apart from this change, the derivations above go through \emph{mutatis-mutandis} in exactly the same way. 

Since the LLL-projected Coulomb interaction has a large $V_0$ component, further varying $V_0$ has little effect on the overlap of the bosonic MR state and the exact ground state. Hence, for bosons, we choose to vary $V_2$ instead. While we do not need an effective interaction for pure LLL Coulomb, we use the following effective interaction for LLL Coulomb with some $\delta V_2$ added to it:
\begin{equation}
V^{\rm eff}(r)={1\over r}+\left(C_0+C_1r^2+C_2r^4\right)e^{-r^2}.
\label{eq: eff_int_LLL_dV2}
\end{equation}
We find the values of coefficients $C_0$, $C_1$ and $C_2$ through matching the $V_0$, $V_2$ and $V_4$ components. The numerical values for the coefficients are listed in Table~\ref{coff2}.
\begin{table}[h]
\begin{tabular}{|c|c|c|c|}
\hline
$\delta V_2$ & $C_0$ & $C_1$ & $C_2$\\ \hline
0.0&0&0&0\\ \hline
0.04&-6.250000&10.937500&-1.953125\\ \hline
0.08&-12.50000&21.875000&-3.906250\\ \hline
\end{tabular}
\caption{\label{coff2}The coefficients of the effective interaction $V^{\rm eff}$ defined in Eq.~\eqref{eq: eff_int_LLL_dV2} that reproduces the same $V_0$, $V_2$ and $V_4$ pseudopotentials as the LLL Coulomb interaction with some $\delta V_2$ added to it. }
\end{table}
The sum of the electron-background and background-background interaction for the LLL Coulomb plus $\delta V_2$ interaction is given by~\cite{Balram20b}
\begin{equation}
 {1\over \sqrt{Q}}+\delta V_1{4Q-3\over (2Q+1)^2},   
\end{equation}
where $2Q{=}N{-}2$ is the flux at which the bosonic $\nu{=}1$ MR state occurs.

The dispersion of the GMP and NF modes at $\nu{=}1$ and finite-size extrapolation of their long-wavelength gaps are shown in Fig.~\ref{bosonic}. For the pure LLL Coulomb interaction, there is a SUSY gap which is about $1/4$ of the GMP gap at $k{\rightarrow} 0$. The SUSY gap decreases with the enhancement of the $V_2$ pseudopotentials. Near $\delta V_2{=}0.08$, the SUSY gap drops to zero. We note that the overlap between the exact ground state and MR state is also maximized around $\delta V_2{=}0.08$ (data not shown). Thus, we conclude that bosons at $\nu{=}1$ behave similarly to fermions at $\nu{=}5/2$, discussed in the main text.

\begin{figure*}
		\includegraphics[width=0.32\textwidth]{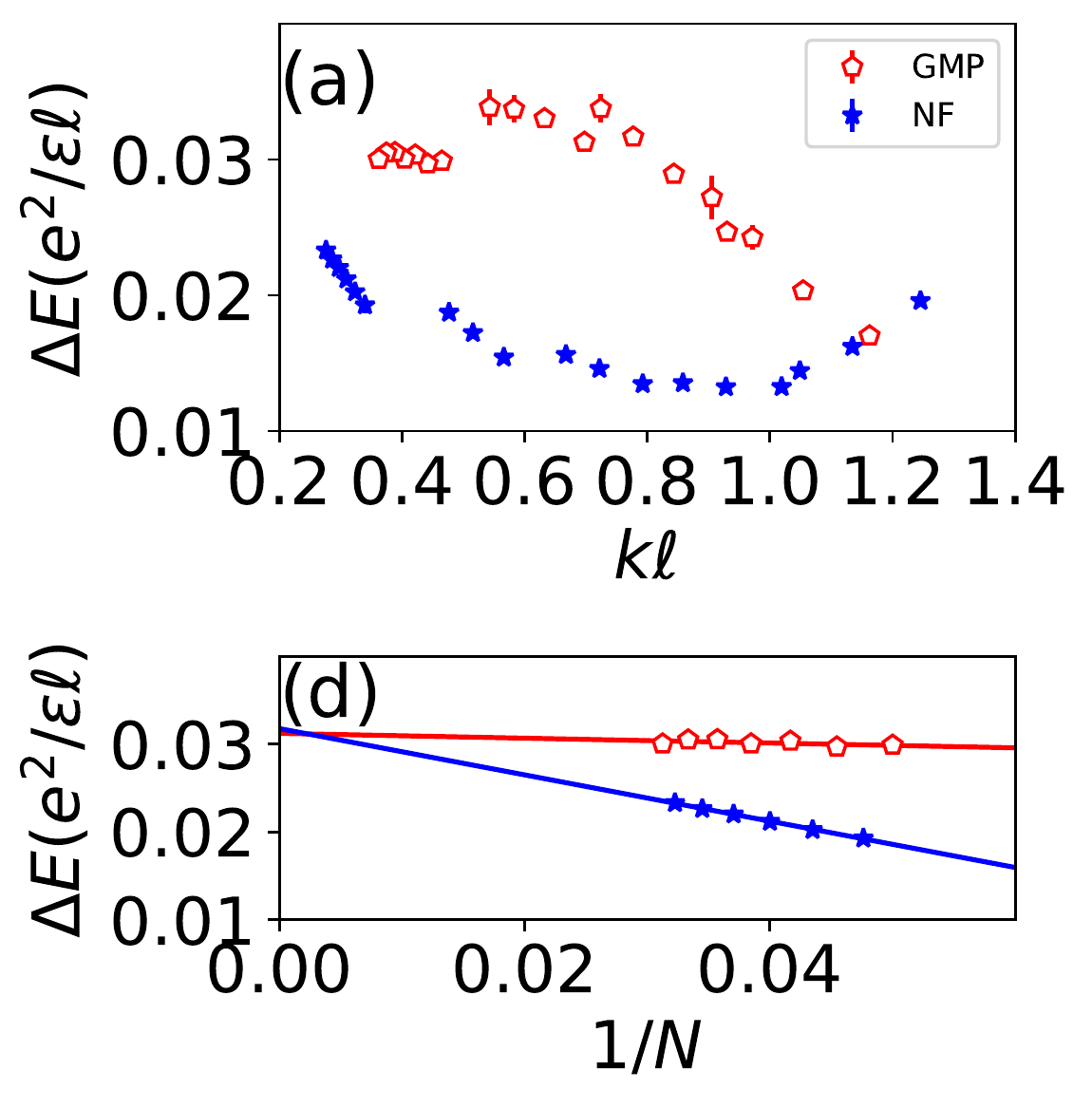}
        \includegraphics[width=0.32\textwidth]{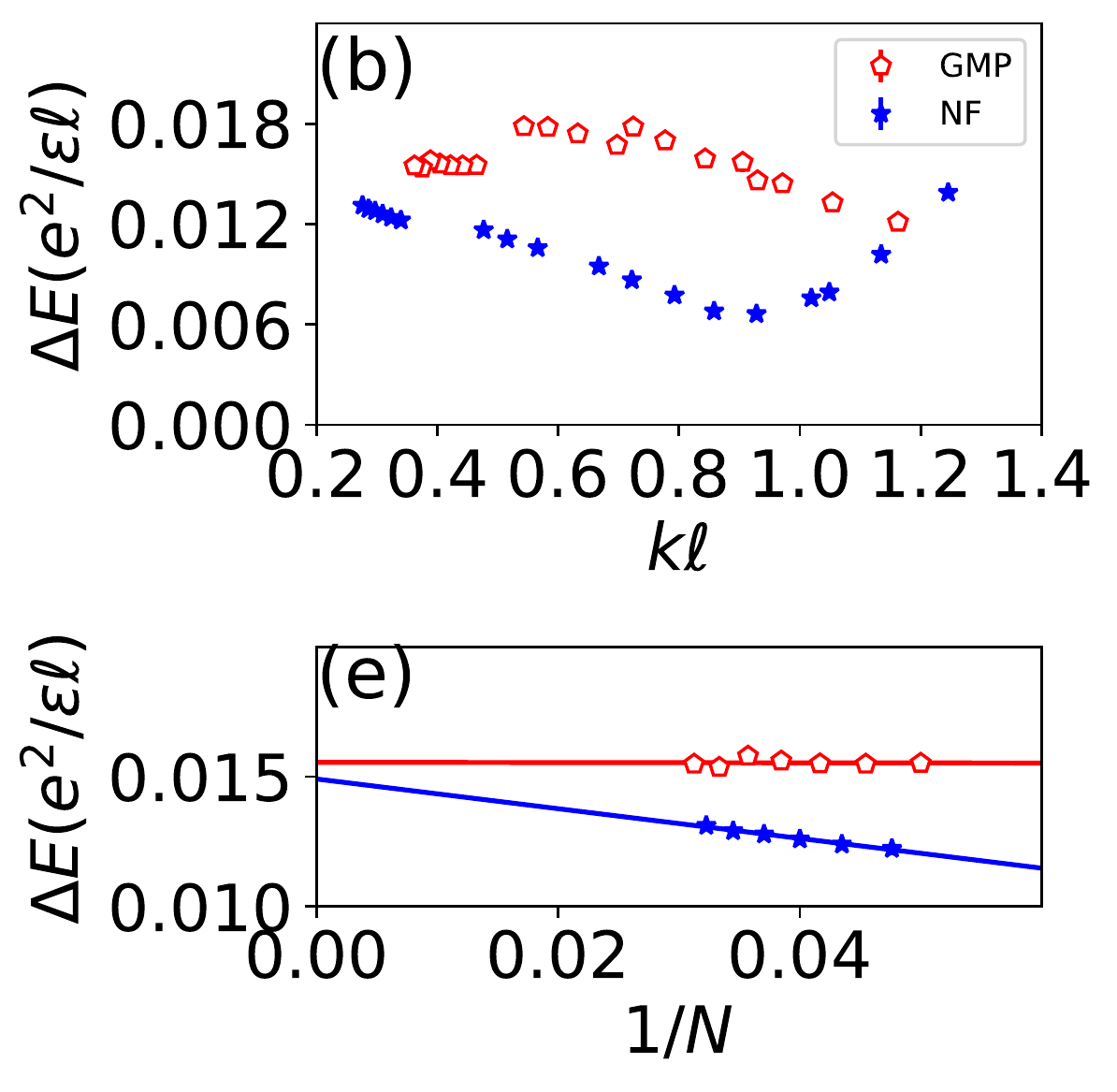}
        \includegraphics[width=0.32\textwidth]{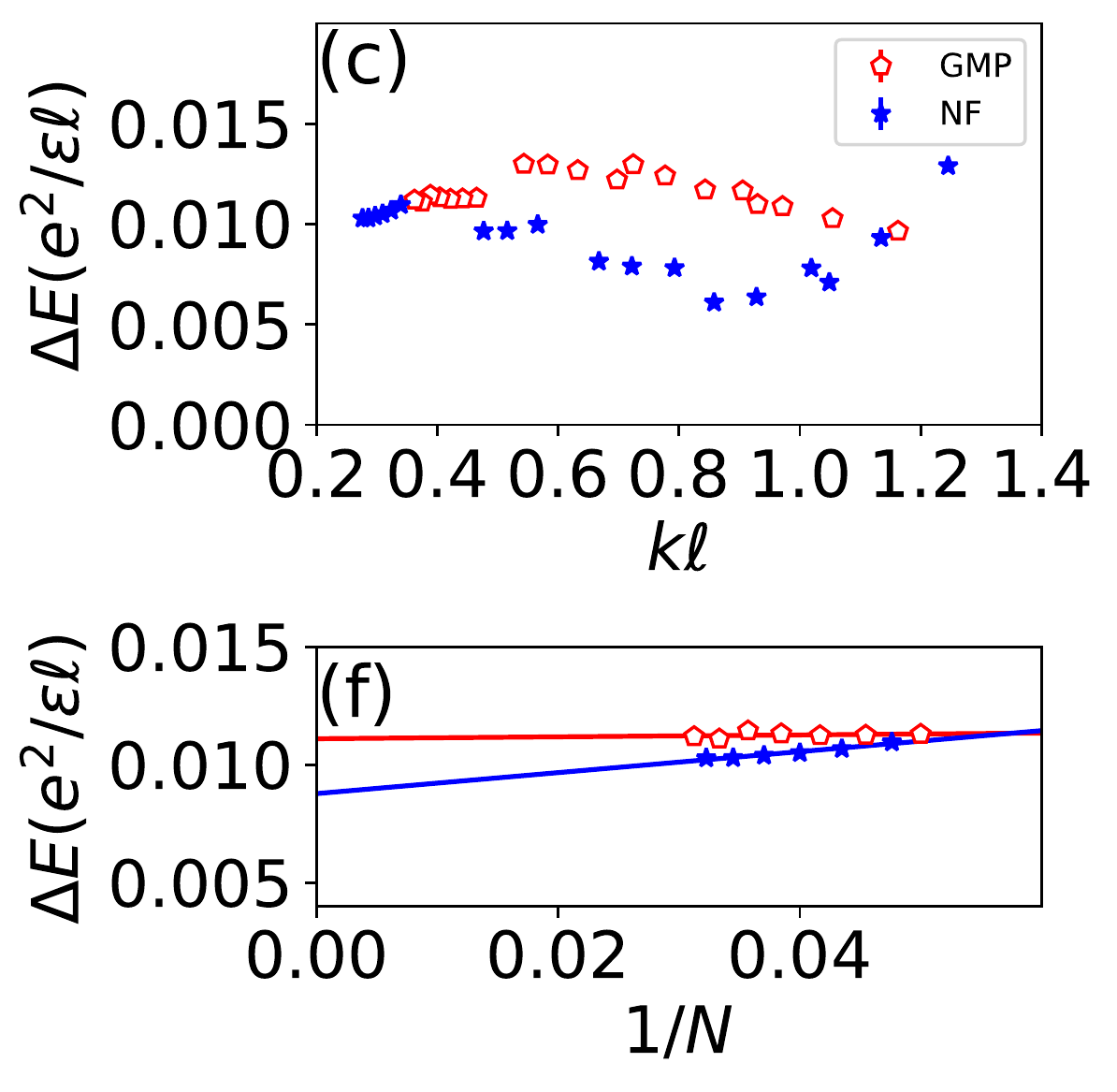}
		\caption{
  Energy dispersion (a)-(c) and finite-size extrapolation (d)-(f) of the GMP and NF mode energies for electrons at $\nu{=}5/2$. The interaction is second-LLL-projected Coulomb with finite width modelled according to $1/\sqrt{r^2{+}w^2}$, where $w{=}\ell$ for (a) and (d), $w{=}2\ell$ for (b) and (e), $w{=}2.5\ell$ for (c) and (f). The energies shown are measured relative to the ground state energy, i.e., $\Delta E{=}E{-}E_0$. The plots include data from various system sizes for $N {\in} [20, 32]$. The finite-size extrapolation uses the energies of $L{=}2$ GMP mode and $L{=}3/2$ NF mode only. All energies are in units of $e^2/\epsilon\ell$.
		}
  \label{width}
\end{figure*}

\section{Effect of finite width}
\label{finite-width}

One of the experimental parameters that can be directly tuned is the finite width of the semiconductor quantum well. We consider the SLL Coulomb interaction with finite width modeled as $1/\sqrt{r^2{+}w^2}$~\cite{Peterson08b}, where $w$ is the width of the quantum well. The effective interaction in the LLL that has the same pseudopotentials as the finite-width interaction in the SLL is
\begin{equation}
\label{finite}
V^{\rm eff}(r)={1\over \sqrt{r^2+w^2}}+{r^2-2w^2\over\left(r^2+w^2\right)^{5/2}}+{3\left(8w^4-24w^2r^2+3r^4\right)\over 4\left(w^2+r^2\right)^{9/2}}.
\end{equation}
The dispersion of the GMP and NF modes evaluated for the effective interaction given in Eq.~\eqref{finite} for various values of $w$, is shown in Fig.~\ref{width}. Comparing these results to the pure second LL Coulomb interaction results shown in the main text, we see that the GMP mode has higher energy than the NF mode for $w{<}\ell$, and nearly the same energy as the NF mode for $\ell{<}w{<}2\ell$, and again higher energy than NF mode when $w$ further increases. We note that the exact ground state has the maximal overlap with the MR state when $w{\sim}2\ell$~\cite{Peterson08b, Papic09}, which is in the range where the SUSY gap is nearly zero.

\section{Effect of $V_3$ Haldane pseudopotential}

In the main text, we probed the phase diagram of the $\nu=5/2$ state in the vicinity of the pure Coulomb interaction, projected to the SLL, by varying the $V_1$ Haldane pseudopotential. To gain a more complete understanding of the phase diagram, one can study the overlap with the Moore-Read state as one simultaneously introduces both $\delta V_1$ and $\delta V_3$ pseudopotential modifications of the Coulomb interaction at $\nu=5/2$. 

\begin{figure}[tbh]
    \centering
    \includegraphics[width=\linewidth]{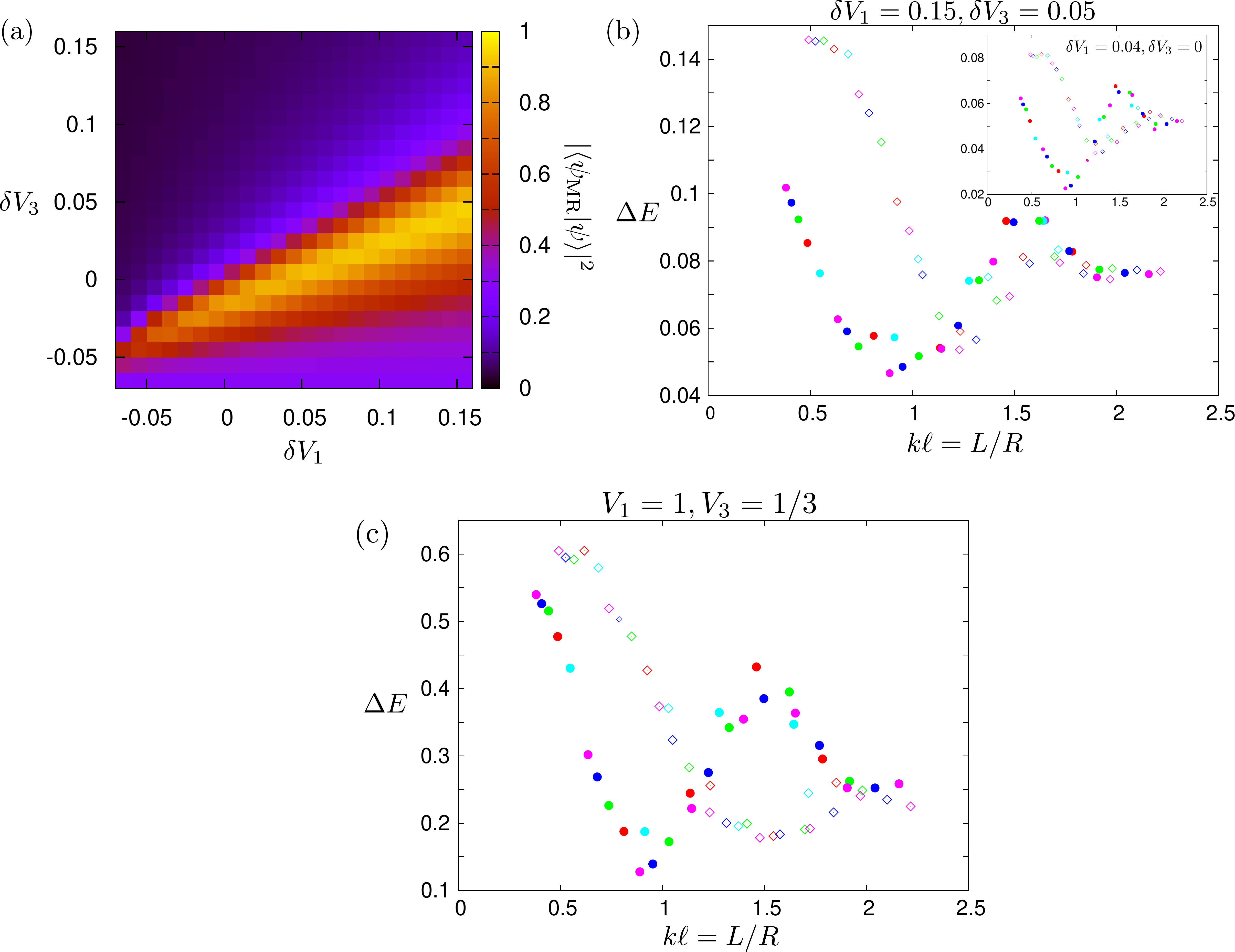}
    \caption{(a) Overlap of the Moore-Read state with the ground state of Coulomb interaction in the SLL, modified by $\delta V_1$ and $\delta V_3$ pseudopotentials. The ground state of the pure Coulomb interaction ($\delta V_1=\delta V_3=0$) is close to the boundary of the region with high overlap. The overlap is further enhanced by simultaneously increasing $\delta V_1$ and $\delta V_3$ in approximately $3:1$ ratio. Data is for $N=14$ electrons. (b) The variational energies of the GMP and NF collective modes, obtained via Jack polynomials in system sizes $N\leq 18$ electrons. The collective mode dispersions are qualitatively similar at large pseudopotential variation $\delta V_1=0.15$, $\delta V_3=0.05$ compared to the case considered in the main text, shown in the inset.
    (c) Collective mode dispersions for the short-range interaction defined by $V_1=1$, $V_3=1/3$~\cite{Peterson08a}, representing the particle-hole symmetrized Moore-Read 3-body interaction (up to small corrections due to the curvature of the sphere). Data is obtained by the same method and for the same system sizes as in (b).
    }
    \label{fig:V3}
\end{figure}

In Ref.~\cite{Rezayi00} it was found that varying only $V_3$ has an approximately opposite effect to varying only $V_1$ pseudopotential: the overlap with the Moore-Read state was found to increase for small variations in either $\delta V_1>0$ or $\delta V_3 < 0$. Varying both pseudopotentials simultaneously, Ref.~\cite{Storni10} found a region in the $\delta V_1-\delta V_3$ phase diagram where the ground state exhibits an enhanced overlap with the Moore-Read state. This region primarily extends in the positive range of $\delta V_1$, $\delta V_3$, i.e., it corresponds to interactions that are effectively shorter-ranged compared to Coulomb. Moreover, as we demonstrate in Fig.~\ref{fig:V3}(a), high overlap is obtained when the added pseudopotentials are approximately in ratio $\delta V_1 \approx 3 \delta V_3$. 
However, the variation of the overlap is small throughout this region. We find the maximal overlap to be $\approx 0.94$, while the maximal overlap achieved by only varying $\delta V_1$ is $\approx 0.93$ (at system size $N=14$). Note that if we keep increasing $\delta V_1$, $\delta V_3$ beyond the scale of Fig.~\ref{fig:V3}(a), the large overlap persists until we reach the limit $\delta V_1, \delta V_3 \gg 1$, which corresponds to the particle-hole symmetrized 3-body interaction of the Moore-Read state~\cite{Peterson08a}.

Given that the ground state does not change much in the high-overlap region of the phase diagram in Fig.~\ref{fig:V3}(a), it is perhaps unsurprising that we also find the collective mode dispersions to be relatively unchanged as we vary $\delta V_1 \approx 3\delta V_3$, see Fig.~\ref{fig:V3}(b) for an illustrative choice of pseudopotential parameters. Compared with the case of varying only $\delta V_1$, considered in the main text and reproduced in the inset of Fig.~\ref{fig:V3}(b), we see little change in the shape of the collective modes (apart from the change in the gap, which is expected due to the difference in overall energy scale). For comparison, in Fig.~\ref{fig:V3}(c) we present the collective mode dispersions for the short-range interaction containing only $V_1=1$ and $V_3=1/3$ pseudopotentials~\cite{Hutzel18}. In this case, we see an even closer convergence of the energies of GMP and NF modes, even at relatively small system sizes that can be accessed via exact diagonalization.   

In conclusion, one might expect to find SUSY degeneracy across the high-overlap region in the phase diagram in Fig.~\ref{fig:V3}(a) and not just at the special line $\delta V_3=0$ studied in the main text. However, the dispersions shown in Fig.~\ref{fig:V3}(b)-(c) were obtained using Jack polynomials~\cite{Yang12b} in relatively small systems, which do not allow a reliable extrapolation to the thermodynamic limit. Unfortunately, we have been unable to extend this calculation to larger systems using the Monte Carlo method presented in the main text. This is due to large error bars in Monte Carlo energy estimates, resulting from the more non-local real-space interaction that effectively describes the $V_3$-modified Coulomb potential. 

\section{Monte Carlo simulation}

In this section, we briefly summarize the key details of the Metropolis Monte Carlo technique used in this work. More details about applying Metropolis Monte Carlo to FQH states can be found in Ref.~\cite{Jain07}. We evaluate the energy as
\begin{equation}
    E=\prod_{i=1}^N\int d\vec{r}_i{|\Psi(\vec{r}_1\cdots\vec{r}_N)|^2\sum_{m<n}V(\vec{r}_m-\vec{r}_n)\over |\Psi(\vec{r}_1\cdots\vec{r}_N)|^2}.
\end{equation}
Here $V(\vec{r}_m-\vec{r}_n)$ stands for the effective interaction introduced in previous sections, see Eqs.~\eqref{eq: eff_int_SLL_dV1} or \eqref{eq: eff_int_LLL_dV2}, $\Psi(\vec{r}_1\cdots\vec{r}_N)$ stands for the LLL projected GMP or NF wave function presented in the main text, and we use the unnormalized $|\Psi(\vec{r}_1\cdots\vec{r}_N)|^2$ as the weight function in the Metropolis sampling process. We evaluate the energy for each angular momentum independently as the weight functions correspond to different angular momentum states. We set the thermalization steps to be $20,000$, and keep the acceptance ratio around $0.5$. For each data point, we run ten independent sampling chains, with over $10^7$ iterations in each chain.

\end{document}